\documentclass [aps,prl,preprint,showpacs,preprintnumbers,amsmath,amssymb,groupedaddress]{revtex4}
\usepackage{graphics}
\usepackage{graphicx}
\usepackage{epsfig}

\newcommand{\beq}{\begin{equation}}
\newcommand{\eeq}{\end{equation}}
\newcommand{\bea}{\begin{eqnarray}}
\newcommand{\eea}{\end{eqnarray}}
\usepackage{dcolumn}
\usepackage{bm}

\begin{document}

\title{Electrical Measurement of the Direct Spin Hall Effect in Fe/In$_{x}$Ga$_{1-x}$As Heterostructures}

\author{E. S. Garlid}
\affiliation{School of Physics and Astronomy, University of Minnesota, Minneapolis, MN 55455}
\author{Q. O. Hu}
\affiliation{Dept. of Electrical and Computer Engineering, University of California, Santa Barbara, CA 93106}
\author{M. K. Chan}
\affiliation{School of Physics and Astronomy, University of Minnesota, Minneapolis, MN 55455}
\author{C. J. Palmstr\o m}
\affiliation{Dept. of Electrical and Computer Engineering, University of California, Santa Barbara, CA 93106}
\affiliation{Dept. of Materials, University of California, Santa Barbara, CA 93106}
\author{P. A. Crowell}
\affiliation{School of Physics and Astronomy, University of Minnesota, Minneapolis, MN 55455}

\begin{abstract}
We report on an all-electrical measurement of the spin Hall effect in epitaxial Fe/In$_{x}$Ga$_{1-x}$As heterostructures with $n$-type channel doping (Si) and highly doped Schottky tunnel barriers.   A transverse spin current generated by an ordinary charge current flowing in the In$_{x}$Ga$_{1-x}$As is detected by measuring the spin accumulation at the edges of the channel. The spin accumulation is identified through the observation of a Hanle effect in the Hall voltage measured by pairs of ferromagnetic contacts. We investigate the bias and temperature dependence of the resulting Hanle signal and determine the skew and side-jump contributions to the total spin Hall conductivity.
\end{abstract}
\maketitle

The generation and manipulation of spin populations in a single device by making use of spin-orbit coupling has been a longstanding goal of the field of semiconductor spintronics \cite{D'yakonov:PLA:1971,Datta:APL:1990}.  There has been extensive theoretical discussion of the spin Hall effect (SHE) and the various ways that it could be exploited to generate or manipulate spin currents \cite{Sinova:PRL:2004,Schliemann:PRB:2004,Engel:PRL:2005,Bernevig:PRB:2005,Tse:PRL:2006,Engel:PRL:2007}.  However, only a handful of recent experiments have investigated this effect, and in semiconductor materials they have relied on optical techniques to either detect \cite{Kato:Science:2004,Wunderlich:PRL:2005,Sih:NatPhys:2005,Stern:PRL:2006,Matsuzaka:PRB:2009} or generate \cite{Wunderlich:NatPhys:2009} spins.   The scope of experimental studies could be broadened significantly by access to transport techniques that can probe both materials and device geometries that are not accessible optically.

\begin{figure}
    \includegraphics*[width=8.5cm]{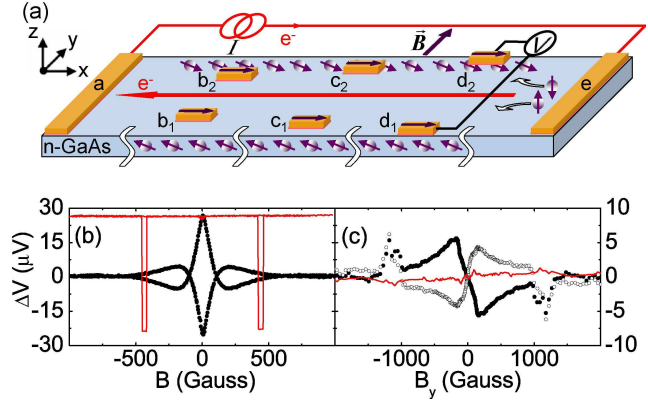}
    \caption{(color on-line) (a) Schematic diagram (not to scale) of the device layout and SHE measurement.  In the actual experiment, contacts at different edge separations are on different devices on the same chip and the current injection contacts are located $> 250~\mu$m away from Hall measurement contacts.  (b) Non-local spin valve ($\relbar$) and Hanle effect ($\bullet$) data obtained on a GaAs device at $T=60$~K for $j_{Inj} = 8.2\times 10^2$~A/cm$^2$. Hanle data are shown for both parallel and antiparallel states of injector and detector. (c) SHE data obtained on a GaAs device at $T=30$~K with the two Hall contacts in the parallel state for $j_x = \pm 5.7 \times 10^{3}$~A/cm$^{2}$ ($\bullet$ and $\circ$) and antiparallel state ($\relbar$).}
    \label{fig:fig1}
\end{figure}

In this Letter we report on an all-electrical measurement of the SHE in lateral devices fabricated from Fe/In$_{x}$Ga$_{1-x}$As heterostructures doped just above the metal-insulator transition.  The SHE is due to spin-orbit scattering of an ordinary charge current, resulting in a transverse spin current.  In the geometry shown in Fig.~\ref{fig:fig1}(a), a charge current $j_{x} = \sigma_{xx}E_{x}$ flows down a channel of conductivity $\sigma_{xx}$  in the presence of an electric field $E_{x}$.  The electrons have a drift momentum $\hbar {\mathbf k} = (m^{*} j_{x}/ne) \hat x$, where $n$ is the carrier density and $m^{*}$ is the effective mass.   The electron spins interact with impurities via the spin-orbit Hamiltonian
$H_{so} = \lambda_{so}{\boldsymbol \sigma}\cdot({\mathbf k}\times \nabla V),$
where $\lambda_{so}$ is the spin-orbit coupling constant, $\boldsymbol \sigma$ is the Pauli spin operator, and  $V$ is the Coulomb-like scattering potential.  The scattering process leads to spin-dependent deflection of electrons, resulting in a spin current $j_{s}$ perpendicular to both their spin orientation and drift momentum.  In steady state, this process leads to an accumulation of spins of opposite sign at the two edges of the channel, and we detect this spin accumulation using Fe contacts as spin-dependent Hall voltage probes.   For the geometry shown in Fig.~\ref{fig:fig1}(a), we are sensitive to a spin current $j_{s}$ that flows in the $y$ direction, with the spin oriented along $z$.   We find that the magnitude of the spin Hall conductivity $\sigma_{SH}=j_{s}/E$ is in agreement with models of the extrinsic SHE due to ionized impurity scattering \cite{Engel:PRL:2005,Tse:PRL:2006}.  By analyzing the dependence of the SH signal on channel conductivity, we determine the relative magnitudes of the skew and side-jump contributions to the total spin Hall conductivity.  We find that the ratio of these terms is approximately constant, independent of the spin orbit coupling parameter, but  the relative magnitude of the side-jump contribution is consistently larger than predicted by theory \cite{Engel:PRL:2005,Tse:PRL:2006}.  The temperature dependence of the spin Hall conductivity is weak over the range of our experiment ($T < 150$~K), although our sensitivity at the highest temperatures is limited by the short spin lifetime in the channel.

Epitaxial (001) Fe/In$_{x}$Ga$_{1-x}A$s heterostructures were grown by molecular beam epitaxy.  The semiconductor epilayers were 2.5~$\mu$m thick and with Si-dopings between $n = 3 \times 10^{16}$~to~$5 \times 10^{16}$~cm$^{-3}$.  Highly doped Schottky tunnel barriers ($n^{+} = 5 \times 10^{18}$~cm$^{-3}$) were prepared as described in Ref.~\onlinecite{Lou:NatPhys:2007}.  Four heterostructures with In concentrations $x =$~0.00, 0.03, 0.05, and 0.06 were studied.  The wafers were subtractively processed into devices using standard lithographic and etching techniques \cite{Lou:NatPhys:2007}.  Multiple devices were fabricated on a single chip with 30$~\mu$m wide channels oriented along the [110] direction, which is the $x$-direction in the schematic diagram of Fig.~\ref{fig:fig1}(a).  Pairs of Fe electrodes, each of which is 4~$\mu$m wide, were patterned so that the centers of the contacts in each pair are 2, 6, or 10$~\mu$m from the edges of the channel. The device geometry is shown schematically in Fig.~\ref{fig:fig1}(a), in which each of the contact separations is illustrated on a single device for simplicity.  Current injection contacts are located at the ends of the channel, $> 250~\mu$m away from the Hall contacts.  The charge current $j_x$ is therefore unpolarized.

Typical spin valve and Hanle effect curves (see Ref.~\onlinecite{Lou:NatPhys:2007} for a discussion) on a lateral spin valve device are shown in Fig.~\ref{fig:fig1}(b).  These data establish that the FM contacts are sensitive to the spin polarization generated by spin injection into the channel as well as its dephasing by precession in an applied magnetic field.  We also establish that the Fe contacts, which have an easy axis along [110], show sharp and reproducible switching behavior as well as nearly perfect remanence.  The hard axis direction is [1$\overline 1$0], which is parallel to the the $y$-axis in Fig.~\ref{fig:fig1}(a).  The hard axis saturation field is $\approx 1.5$~kOe.

Since the contacts are magnetized along [110] ($\hat x$), and the spin polarization generated by the spin Hall effect is expected to be oriented along [001] ($\hat z$), a field $B_y$ is applied to precess the spin accumulation into the [110] direction \cite{Engel:PRL:2007,Crooker:Science:2005}.  We therefore expect to observe an increase in the spin Hall voltage at low fields followed by a suppression due to spin dephasing in large fields.  The signal should reverse sign when $B_y$ is reversed.  Although the spin accumulation is small (a few $\mu$V), we will show that it can be identified by the expected dependence on $B_y$.  In practice, however, there are three significant background effects that obscure the spin Hall signal: 1) the ordinary Hall effect due to the applied field, 2) local Hall effects due to the fringe fields generated by the FM contacts; and 3) voltages due to the small fraction (0.1\%) of the channel current that is shunted through the Hall contacts.  The first two backgrounds can be eliminated fairly easily based on expected symmetries. For example, reversing the magnetizations of both Hall contacts from the $+ x$ to $- x$ directions does not change the ordinary Hall voltage but reverses the sign of the spin Hall voltage.  By taking the difference of two field sweeps with the Hall contacts in the two different parallel states, the ordinary Hall effect is thus removed.  Local Hall effects are due predominantly to the $x$-components of the contact magnetization.  The corresponding fringe fields, which are in the $\pm z$-direction are {\it even} with respect to $B_y$, while the spin Hall voltage is odd in $B_y$.  We can therefore eliminate local Hall effects by retaining only the components of the signal that are odd with respect to $B_y$.

Data taken on a GaAs sample with a channel current $j_x = \pm 5.7 \times 10^{3}$~A/cm$^{2}$ at $T=30$~K are shown in Fig.~\ref{fig:fig1}(c) after removing the first two background contributions. By construction, these data are odd with respect to $B_y$, and they show extrema at intermediate fields (approximately 250 Oe) as expected.  Although the exact form of these data will be discussed below, the magnitude of the Hall voltage at these maxima corresponds to a spin polarization  $P = (n^\uparrow - n_\downarrow)/(n^\uparrow + n_\downarrow)\approx 1.3\%$ at the sample edges, where
\begin{eqnarray}
P = \frac{e \Delta V}{\eta P_{Fe}}\frac{3 m^{*}}{\hbar^{2}(3\pi^{2}n)^{2/3}}.
\label{eq:PfromV}
\end{eqnarray}
In this expression, which follows from the usual relationship between the spin accumulation and the density of states \cite{Lou:NatPhys:2007}, $P_{Fe}=0.42$ is the spin polarization of Fe at the Fermi level and $\eta \approx 0.5$ is the interfacial transparency.
There are, however, additional features in the field sweeps near 1~kOe that do not reverse sign when the current is reversed, and hence cannot be due to a Hall effect.  These result from the current that is shunted through the Fe contacts (and hence has a component perpendicular to the plane) in combination with tunneling anisotropic magnetoresistance (TAMR) at the Schottky contact \cite{Moser:PRL:2007}.  This final background contribution can be minimized by subtracting the Hall voltage for the two current directions, as will be done for all subsequent data shown in this paper.

We have also performed  the same measurements with the FM contacts on opposite sides of the channel initialized in either of the antiparallel states $\uparrow\downarrow$ and $\downarrow\uparrow$.  The data in this case are shown as the solid line in Fig. \ref{fig:fig1}(c) after removal of all three backgrounds.  This curve shows no Hall signal, demonstrating that the spin accumulations at opposite edges of the sample are opposite in sign.

\begin{figure}
    \includegraphics*[width=8.5cm]{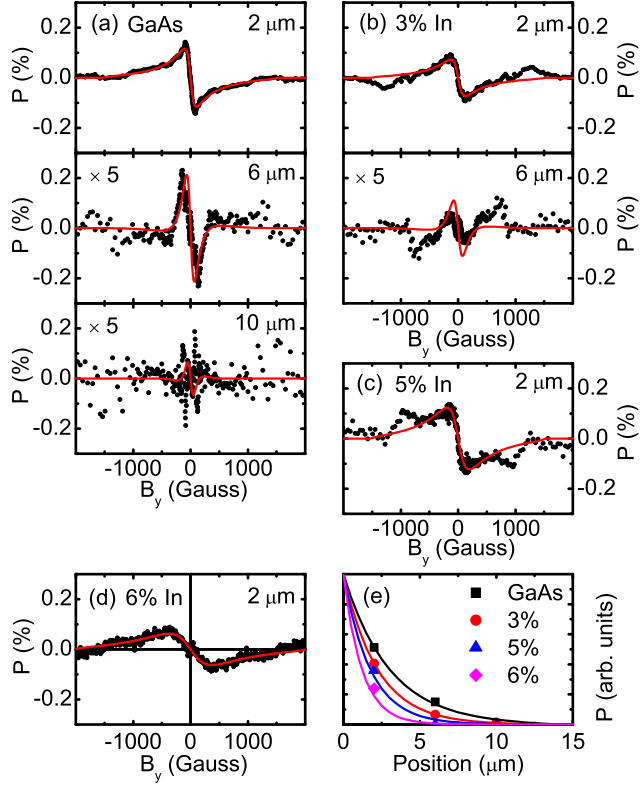}
    \caption{(color on-line) (a) SHE signal (spin polarization $P$ deduced from the Hall voltage $\Delta V$) ($\bullet$) as a function of magnetic field for the GaAs sample at $T = 30$~K and channel current $j_{x} = \pm 2.9 \times 10^{3}$~A/cm$^{2}$.  The solid curves ($\relbar$) show fits to all contact separations with a single set of parameters.  The data obtained at 6 and 10$~\mu$m from the edges are multiplied by 5.  (b) Data and fits for the In$_{0.03}$Ga$_{0.97}$As sample under the same bias conditions.  No spin signal is resolved for contacts 10$~\mu$m from the edge.  (c) Data and fit for the In$_{0.05}$Ga$_{0.95}$As sample under the same conditions.  No spin signal is resolved for contacts 6 and 10$~\mu$m from the edge.  (d) Data and fit for the In$_{0.06}$Ga$_{0.94}$As sample under the same conditions.  (e) Magnitude of the SHE signal as a function of distance from the edge of the channel for all four samples, normalized to $P_0$.  Solid lines show decay of spin polarization with distance as determined from fit parameters.}
    \label{fig:fig2}
\end{figure}

Data taken at different contact separations for the $x =$~0, 0.03, 0.05 and 0.06 devices at $T=30$~K and $j_{x} = \pm 2.9\times 10^3$~A/cm$^2$ are shown in Fig. \ref{fig:fig2}.  The spin Hall voltage $\Delta V$  has been converted to spin polarization using Eq.~\ref{eq:PfromV}. The polarization at contacts further from the edges of the channel shows a field dependence that is qualitatively similar to that observed at the edges, but with a smaller magnitude and narrower width.  The devices with a non-zero In concentration $x$ show a smaller spin Hall signal that decays more rapidly with distance from the edges of the channel.  No spin signal is observed 10$~\mu$m from the edge for $x=0.03$ or at 6 and 10$~\mu$m from the edge for $x = 0.05$ and $0.06$. As confirmed by non-local measurements, these samples have shorter spin diffusion lengths than the GaAs sample.  Data similar to those shown in Fig.~\ref{fig:fig2} were taken over a bias range of $j_{x} = 0$~to~$ \pm 5.7 \times 10^{3}$~A/cm$^{2}$ at $T = 30$~K and a temperature range of $T = 30$~to~$200$~K at $j_{x} = \pm 5.7 \times 10^{3}$~A/cm$^{2}$.

We now use these data to determine the magnitude and sign of the spin Hall conductivity.  The first step in this process is to determine the transverse spin current $j_{s}\hat y$, which can be related to the steady-state spin polarization $P_{0}$ at the channel edges by the diffusion equation, so that $j_s = e P_0 n L_{s}/\tau_{s},$ where $\tau_s$ is the spin relaxation time and $L_s = \sqrt{D\tau_s}$ is the spin diffusion length.  Determining $P_0$ requires a full fit of $P(B_y)$ to a model that includes precession, diffusion, and spin relaxation.  This is essentially identical to the usual analysis of non-local Hanle measurements in a lateral spin valve \cite{Lou:NatPhys:2007} after accounting for the perpendicular orientation of the ``source'' (the spin Hall current) with respect to the detector.  To constrain the fits, the diffusion constant $D$ is obtained from the channel conductivity $\sigma_{xx}$ and the carrier density $n$ using the Einstein relation \cite{Flatte:PRL:2000}, and the $g$-factor for each sample is fixed using the value for GaAs, $g = -0.44$ and the dependence on $x$ determined from the $8 \times 8$ Kane model \cite{Silva:PRB:1997,Kiselev:PRB:1998,Madelung:1996,Kato:Science:2004}.  The modeling also includes the rotation of the magnetization by the applied field.  This leaves $P_0$ and $\tau_s$ as the only fitting parameters.  For each bias current, a single set of parameters is used to fit the data sets obtained at different distances from the edge of the channel.  The fitting results are shown as solid curves in the seven panels of Fig.~\ref{fig:fig2}(a-d), and $P$ as a function of position for all four samples is shown in Fig.~\ref{fig:fig2}(e).  The principal features of the data are captured by the fitting, including the decrease with separation from the edges and the location of the extrema, which shift towards smaller field as the Hall contacts are moved towards the center of the channel.  Despite the increased g-factor, the curves for the In$_{x}$Ga$_{1-x}$As samples are broader than for the GaAs sample.  This reflects the shorter spin relaxation as $x$ increases, as verified by spin injection measurements.

From the values of $P_{0}$ and $\tau_{s}$ determined from these fits, it is possible to determine $j_{s}$ and the spin Hall conductivity $\sigma_{SH} = j_{s}/E_{x}$.  For the GaAs sample, we find $\sigma_{SH}~\approx~3.0~\Omega^{-1}m^{-1}$, which is of the same order of magnitude as has been estimated from Kerr microscopy measurements \cite{Kato:Science:2004,Matsuzaka:PRB:2009} and is of the same order and sign as has been predicted by theory \cite{Engel:PRL:2005,Tse:PRL:2006}.  The experimental sign is determined using the known orientation of the electrode magnetizations.  To make a more extensive comparison, we consider the result of Engel and co-workers \cite{Engel:PRL:2005}:
\begin{eqnarray}
\sigma_{SH} \approx \frac{2\lambda_{so}}{(a^{*}_{B})^{2}}\sigma_{xx}-\frac{2n\lambda_{so}e^{2}}{\hbar},
\label{eq:SHcondTheory}
\end{eqnarray}
where $\lambda_{so}$ is the spin-orbit coupling parameter, $a^{*}_{B}$ is the effective Bohr radius of an ionized impurity (the presumed source of scattering), and $\sigma_{xx}$ is the channel conductivity.  For the case of GaAs, we use $\lambda_{so} = 5.3$~\AA$^2$~and $a^{*}_{B} = 103$~\AA~(for Si donors) \cite{Engel:PRL:2005}.  At the value of $j_{x}$ used for the data in Fig.~\ref{fig:fig2}, $\sigma_{xx} = 3600$~$\Omega^{-1}m^{-1}$, and Eq.~\ref{eq:SHcondTheory} thus gives $\sigma_{SH} = 2.4$~$\Omega^{-1}m^{-1}$, a factor of 20\% smaller than experiment.  A similar approach was used previously in Ref.~\onlinecite{Matsuzaka:PRB:2009} to analyze the dependence of $\sigma_{SH}$ on doping concentration in GaAs. 

\begin{figure}
    \includegraphics*[width=8.5cm]{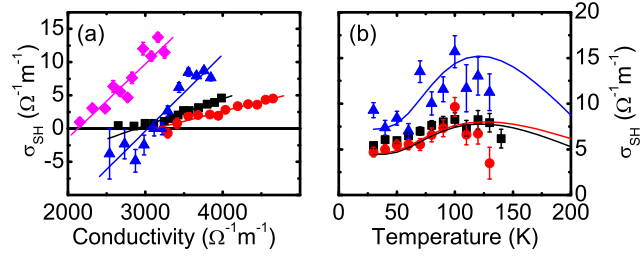}
    \caption{(color on-line) (a) Dependence of $\sigma_{SH}$ on $\sigma_{xx}$ at $T = 30$~K for the $x = $~0 (squares), 0.03 (circles), 0.05 (triangles), and 0.06 (diamonds) samples. A linear fit is used to determine $\gamma$ and $\sigma_{SJ}$ for each sample. The fitting parameters are compiled in Table~I.  (b) Temperature dependence of $\sigma_{SH}$ from $T = 30$~to~$130$~K (solid points) and predicted temperature dependence (curves).}
    \label{fig:fig3}
\end{figure}

As can be seen from examination of Eq.~\ref{eq:SHcondTheory}, there are two expected contributions to the spin-Hall conductivity, one of which scales with $\sigma_{xx}$ (skew scattering) and a second which is a constant of opposite sign (side-jump).  It is possible to tune the mobility, and hence $\sigma_{xx}$, by approximately 25\% by varying the bias current \cite{Oliver:PR:1962}.  We find clear evidence for both contributions in the observed dependence of $\sigma_{SH}$ on $\sigma_{xx}$, which is shown in Fig.~\ref{fig:fig3}(a).  The solid lines in this figure are linear fits.  A negative intercept, indicating the correct sign of the side-jump term, is found in all four samples.  The slopes and intercepts of the fits are given in Table~I.  For the purposes of comparison with theory, we write the total spin Hall conductivity as $\sigma_{SH} = \sigma_{SS} + \sigma_{SJ} = \gamma\sigma_{xx} + \sigma_{SJ}$, where $\gamma$ and $\sigma_{SJ}$ are obtained from Eq.~\ref{eq:SHcondTheory}.   The skewness parameter $\gamma$ for GaAs is about four times larger than the prediction of $\sim 1 \times 10^{-3}$ \cite{Engel:PRL:2005}.  We can also compare our experimental results with the expected ratio $\sigma_{SJ}/\gamma = -(a^{*}_{B})^2 n e^2 \hbar$, which is independent of the spin-orbit coupling $\lambda_{so}$.  We note that there are large systematic errors (particularly those originating from the assumption of a Pauli-like density of states and a fixed $\eta = 0.5$) which appear in the conversion from a Hall voltage to a polarization (Eq.~\ref{eq:PfromV}) and then to a spin current.  These, however, do not impact the ratio $\sigma_{SJ}/\gamma,$ which is about a factor of 2.5 larger than the expected value for all four samples, as shown in Table~I.  
  
\begin{table}
    \centering
    \caption{Fit parameters for $\sigma_{SH}$ vs. $\sigma_{xx}$.}
        \begin{tabular}{|c|cccc|}
    \hline
$\sigma_{SH} = \gamma\sigma_{xx} + \sigma_{SJ}$  &  ~$0\%$~In~   &   ~$3\%$~In~  & ~$5\%$~In~ & $6\%$~In \\
    \hline
$\gamma$   & 0.004 & 0.003 & 0.012 & 0.011 \\
$\sigma_{SJ}~(\Omega^{-1}m^{-1})$   & -12 & -9 & -35 & -28 \\
Meas. $\sigma_{SJ}/\gamma~(10^3~\Omega^{-1}m^{-1})$   & -3.0 & -3.0 & -2.9 & -2.2 \\
Pred. $\sigma_{SJ}/\gamma~(10^3~\Omega^{-1}m^{-1})$ & -1.1 & -1.4 & -1.2 & 0.8 \\
    \hline
    \end{tabular}
\end{table}

We studied the temperature dependence of the SHE in the $x =$~0, 0.03, and 0.05 samples over the range $T = 30$~to~$150$~K at $j_{x} = 5.7 \times 10^{3}$~A/cm$^{2}$.  Figure \ref{fig:fig3}(b) shows the experimentally determined spin Hall conductivity as a function of temperature (solid points).  The solid lines show the predicted temperature dependence of $\sigma_{SH}$ using the values of $\gamma$ and $\sigma_{SJ}$ determined from Fig. \ref{fig:fig3}(a) and the measured values of $\sigma_{xx}(T)$ and $n(T)$.  We find that $\sigma_{SH}$ shows a modest increase over the temperature range of $T = 30$~to~$100$~K due to the increase in electron mobility.  At temperatures above $\approx 120$~K the measured spin Hall polarization decreases rapidly due to the rapid decrease of $\tau_{s}$ with increasing temperature.  This suppresses the spin Hall signal even if $\sigma_{SH}$ is relatively constant.  These data suggest that the spin Hall conductivity is relatively insensitive to phonon scattering, although it is uncertain whether the extrapolated values above $\approx 120$~K are accurate.  The weaker temperature dependence of $\tau_s$ in other materials, such as ZnSe, makes the SHE more readily observable at high temperatures \cite{Stern:PRL:2006}.

The measurements and analysis presented here conclusively demonstrate electrical detection of the direct SHE in Fe/In$_{x}$Ga$_{1-x}$As heterostructures.  The bias and temperature dependences of the SHE indicate that both skew and side-jump scattering contribute to the total spin Hall conductivity.  The ratios of the side-jump to skew scattering contributions for the four samples are similar but larger than predicted for ionized impurity scattering alone.  Although the spin Hall conductivity increases with In concentration, this cannot be attributed unambiguously to an increase in the spin orbit coupling.  We note, however, that the spin accumulation due to the SHE observed for the higher In concentrations is comparable to that generated by direct spin injection from a FM.  This suggests that the SHE could function as a tool for probing spin-dependent phenomena in materials with large spin-orbit coupling and short spin diffusion lengths.

We acknowledge helpful discussions with M.~E.~Flatt\' e.  This work was supported by the ONR MURI program and NSF Grant No. DMR-0804244, and in part by the NSF MRSEC program under Grant No. DMR-0819885.


\end{document}